# Nanoscale resolution scanning thermal microscopy with thermally conductive nanowire probes.

Maria Timofeeva[a], Alexey Bolshakov[a], Peter D. Tovee[b], Dagou A. Zeze[c], Vladimir G. Dubrovskii[a] and Oleg V. Kolosov[b*]

[a] *Laboratory of Physics of Nanostructures, Nanotechnology Centre, Saint-Petersburg Physics And Technology Centre For Research And Education Of Russian Academy Of Sciences, 8, bld. 3 Khlopina, St. Petersburg, 194021, Russia*
[b] *Physics Department, Lancaster University, Lancaster, LA1 4YB, UK*
[c] *School of Engineering and Computing Sciences, Durham University, Durham DH1 3LE, United Kingdom*

*Corresponding author: Tel:+44-1524593619, E-mail:o.kolosov@lancaster.ac.uk; WWW: www.nano-science.com*

**Abstract** Scanning thermal microscopy (SThM) – a type of scanning probe microscopy that allows mapping thermal transport and temperatures in nanoscale devices, is becoming a key approach that may help to resolve heat dissipation problems in modern processors and develop new thermoelectric materials. Unfortunately, performance of current SThM implementations in measurement of high thermal conductivity materials continues to me limited. The reason for these limitations is two-fold - first, SThM measurements of high thermal conductivity materials need adequate high thermal conductivity of the probe apex, and secondly, the quality of thermal contact between the probe and the sample becomes strongly affected by the nanoscale surface corrugations of the studied sample. In this paper we develop analytical models of the SThM approach that can tackle these complex problems – by exploring high thermal conductivity nanowires as a tip apex, and exploring contact resistance between the SThM probe and studied surface, the latter becoming particularly important when both tip and surface have high thermal conductivities. We develop analytical model supported by the finite element analysis simulations and by the experimental tests of SThM prototype using carbon nanotube (CNT) at the tip apex as a heat conducting nanowire. These results elucidate vital relationships between the performance of the probe in SThM from one side and thermal conductivity, geometry of the probe and its components from the other, providing pathway for overcoming current limitations of SThM.

## 1. Introduction

Modern materials science and technology is increasingly devoted to the control of matter on the nanoscale, with local thermal properties playing major role in the diverse materials used in renewable energy generation (thermoelectrics, photovoltaics), structural composites and in optical and electronic devices [1-6]. In semiconductor processors, the inability to dissipate increasing power density lead to the first aspect of Moore's law to fail because of nanoscale thermal management problems [7-9]. In order to address these needs, the tools are needed that are able to perform thermal measurements in solid state materials on the nanoscale. Unfortunately, most thermal measurement systems are based on optical methods, such as IR thermal emission, Raman spectroscopy or photoreflectance with the spatial resolution limited to 500 nm or greater [10-12]. The promising technique for nanoscale thermal measurements is Scanning Thermal Microscopy (SThM) [13-18], that while showing good performance in studies of polymeric and organic materials, has limited ability to differentiate between high thermal conductivity materials such as used in the semiconductor industry, thermoelectrics or optical devices [19, 20]. The main reasons are – spatial resolution on the range of few tens of nanometres remains well below other scanning probe microscopy (SPM) approaches, low sensitivity to thermal properties of materials of high thermal conductance, worsened by the unstable and weak thermal contact between the thermal sensor and the studied object.

One of possible solutions proposed elsewhere [4, 5, 21] is to use a high thermal conductivity element, such as carbon nanotubes (CNT) or similar acting as a thermal link between the sensor and the sample [22, 23] that will also define the nanometre scale probed area. First experimental tests show the high potential of such approach. In this paper we focus on the understanding of the physical principles underlying the operating envelope of such high performance SThM probes, linking the geometry of the probe and parameters of materials used. We developed an analytical thermal model of the probe and tested its validity via finite elements analysis as well as experimental measurements that allowed us to propose the optimal geometry and materials for such high performance probe including semiconductor nanowires (NW) that may add new functionalities to SThM measurements. We also compare a thermal sensitivity of the probe with nanowire with experimental studies of novel SThM probe with CNT tip, and explore future direction that optimize performance of such SThM probes in air and vacuum environments.

## 2. Theory and simulation.

*2.1 Analytical model of the SThM probe*

Measurement of local thermal properties of a sample could be done with atomic force microscope equipped with SThM probe. Fig. 1 demonstrates the schematic image of widely used SThM probe [13, 14]. The SThM cantilever consists of three zones: first is the largest zone that contains the cantilever base with Au pads, second zone is end of the cantilever contains the heating resistor that also acts as a thermal sensor and apex zone of the probe that is in direct thermal contact with the sample, that is either the same material as a cantilever, e.g. $Si_3N_4$, or can be a NW or CNT [4].

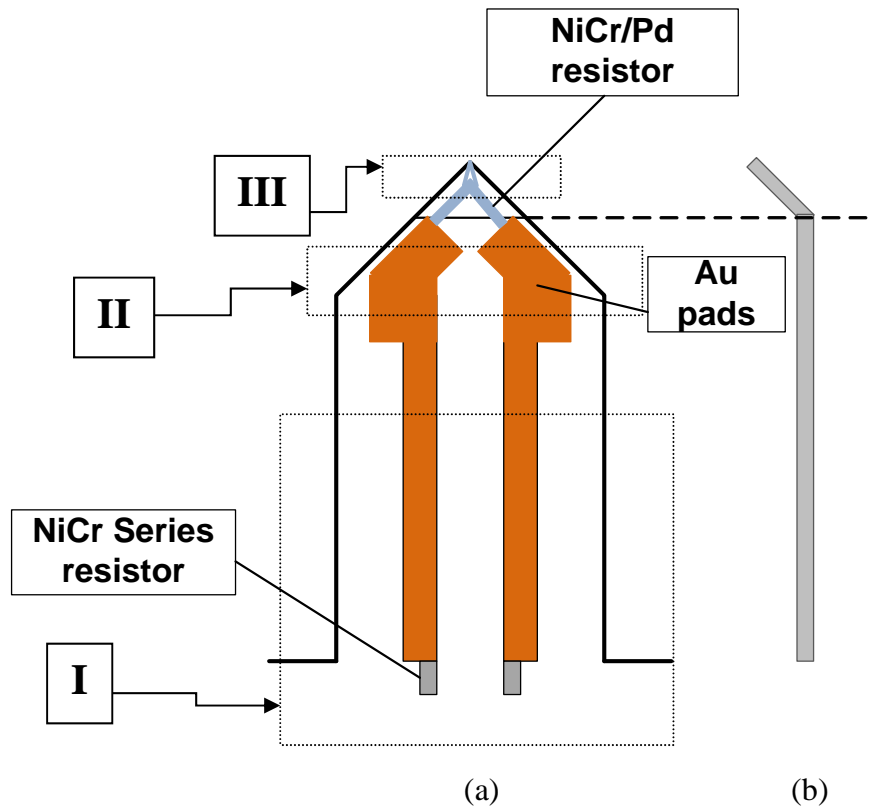

Fig. 1 Schematic image of the SThM cantilever, (I) cantilever base with Au pads, (II) end of cantilever base, (III) SThM tip, (a) – top view, (b) – side view.

In previous publications it was demonstrated that the thermal properties of the tip apex at the end of the cantilever have leading impact on the performance of the SThM probe [4, 6, 24-26]. Therefore in this study we are focusing on the apex parts I and II (Fig.1) of the SThM probe.

The equivalent thermal resistance of the SThM probe is schematically presented in Fig. 2 similar to previously reported models [4, 24].

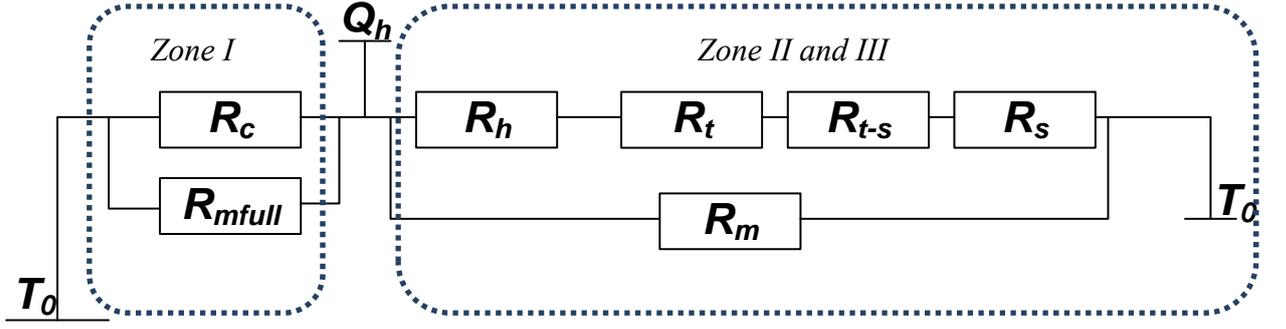

Fig. 2 Thermal resistance diagram of SThM system, including properties of the cantilever, tip of the probe, tip-sample interface, sample and media.

Here $R_c$ – thermal resistance of the cantilever to the holder, $R_{mfull}$ – thermal resistance of cantilever to the environment surrounding cantilever, (zone I) $Q_h$ – heat generated in the heater, $R_h$ – thermal resistance between the heater and the cantilever (zone II), $R_t$ – thermal resistance of the tip apex (zone III), $R_{t-s}$ – thermal resistance of the interface between tip and sample, $R_s$ – thermal (spreading) resistance of heat flow to the sample, $R_m$ – thermal resistance to the environment surrounding end of cantilever and tip (zones II and III), $T_0$ – environment temperature. The heat flow to sample in zones II and III can be expressed as function of these thermal resistances [24]:

$$\frac{Q_h}{T_h - T_0} = \frac{1}{R_m} + \frac{1}{R_h + R_t + R_{t-s} + R_s} \qquad (1)$$

It is clear from this formula that decreasing the tip and contact resistances, and increasing the resistance to the environment, improves the sensitivity of SThM to the sample thermal resistance. That means that we need to decrease both tip and contact resistances as much as possible to increase precision of the SThM measurements.

In the conventional SThM probe, several factors can reduce quality of the measurements. Firstly, it is a relatively large dimension of a tip (with typical radius about 50-100 nm) that leads to a low spatial resolution. Secondly, interface effects, such as contact between two materials and Kapitza resistance of the interface [20] can deteriorate the SThM performance. Finally, the influence of environment could change the effective radius of contact between tip and sample, as well as increase the heat loss to the environment. In this work we analyze the contribution of all these factors to the sensitivity of SThM measurements.

Due to the Au metal leads that reach as far as the base of the tip (see Fig. 1), the tip can be thought of as thermally anchored to the ambient with the resistive heater evenly distributed along the length of the probe. Given that the heat mainly flows down the probe and the tip apex is at first approximation, axially symmetrical cone, it was reasonable to create equivalent axially symmetrical conical model for the cantilever and tip.

In order to analyze a relative contribution of all these factors, we developed an analytical model of key components of SThM cantilever (zones II and III, Fig. 1) that includes a cylindrical thermal conductive tip, such as NW or CNT. Analytical model allows us to investigate influences of tip geometry to the thermal properties of SThM measurements system. Fig. 3 demonstrate two types of generic geometries, first – *"contact"* when tip is attached to the cantilever apex represented by the cylinder of length $L_c$, second type – *"embedded"* tip is attached to the cantilever apex over the length $L_c$. First *"contact"* type correspond to the case when tip is grown directly at the end of cantilever, second *"embedded"* type - when tip is attached to cantilever, for example, CNT (single or multiwall) or NW [25] attached to the end part of $Si_3N_4$ cantilever [5, 27].

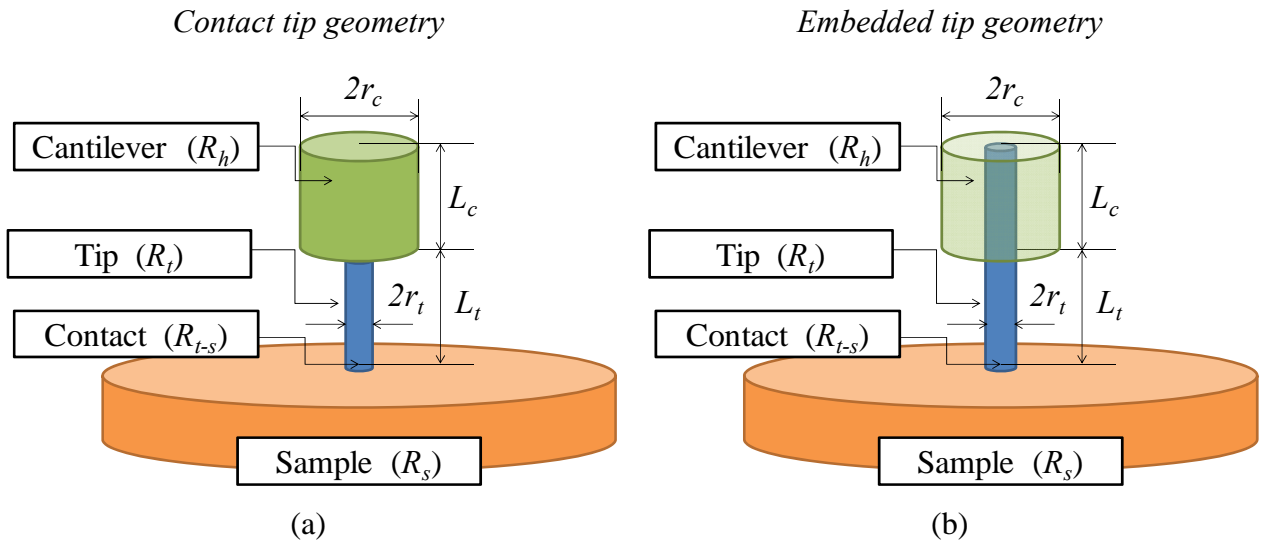

Fig. 3 Schematic images of geometry types, (a) *contact* geometry – tip is in contact with cantilever, (b) *embedded* tip geometry – tip is embedded into cantilever base.

In this model we place the heater at the outer side of the cylinder. In contact geometry the thermal resistance between the cantilever end and the probe ($R_h$) is then the function of $k_c$ – thermal conductivity of cantilever material and contact area between cantilever and tip depending on tip radius ($r_t$). Thermal resistance $R_h$ in the *contact* tip geometry is expressed as [28]:

$$R_h = \frac{1}{2\pi k_c r_t} \quad (2a)$$

Thermal resistance of the cantilever in *embedded* tip geometry can be approximated by thermal resistance of cylinder where $r_c$ – is the outer radius and $r_t$ – the inner radius which is equal to the radius of the probe tip:

$$R_h = \frac{\ln(r_c/r_t)}{2\pi k_c L_c} \quad (2b)$$

Thermal resistance of the tip itself is then:

$$R_t = \frac{L_t}{\pi k_t r_t^2} \quad (3)$$

Here $L_t$ – length of tip, $k_t$ - thermal conductivity of tip material. Contact thermal resistance is also known as Kapitza resistance [29] – is the resistance due to the presence of the interface between two different materials. This interfacial thermal resistance plays a significant role in thermal transport at nanoscale measurements [30]. In addition, when the size of the contact approaches the length of the mean free path of the energy carriers, phonons or electrons, that can range from few nm in amorphous oxides to few tens of nm in metals and few hundreds of nm in Si and graphene, this can further limit the transport energy in nanostructures [31]. In brief, the contact (Kapitsa) component of thermal resistance depends on materials properties and size of contact. Contact thermal resistance could be presented in the form:

$$R_{t-s} = \frac{\rho_{t-s}}{\pi r_t^2} \quad (4)$$

Here $\rho_{t-s}$ – contact thermal resistivity which in this approximation of contact between non-metallic materials depends on the ratio of Debye temperatures [32].

Finally, the spreading thermal resistance of the sample can be presented in the form [28]:

$$R_s = \frac{1}{2\pi k_s r_t} \quad (5)$$

Summarizing all thermal resistances (2a-5) we obtain full thermal resistance for contact-tip geometry:

$$R_{i-m\_con} = \frac{1}{2k_c \pi r_t} + \frac{L_t}{\pi k_t r_t^2} + \frac{\rho_{t-s}}{\pi r_t^2} + \frac{1}{2\pi k_s r_t} \tag{6a}$$

Considering the *"embedded"* tip geometry this thermal resistance is modified as:

$$R_{i-m\_emb} = \frac{\ln(r_c / r_t)}{2\pi k_c L_c} + \frac{L_t}{\pi k_t r_t^2} + \frac{\rho_{t-s}}{\pi r_t^2} + \frac{1}{2\pi k_s r_t} \tag{6b}$$

here the main difference is due to cylindrical heat transfer between the heater and NW via cantilever thickness [33] producing the logarithmic term in the Eq. 6(b). As we see in the section *4.2* below, these values compare reasonably well with the FE simulation and experimental measurements in the SThM. As they provide the explicit dependence of the SThM output as a function of the probe geometry ($r_t$, $L_t$, $L_c$ and thermal properties of the probe and the sample $k_c$, $k_t$, $\rho_{t-s}$) the Eqs. 6a and 6b allow one to highlight the trends of this response for various implementation of NW SThM probes that will be discussed more in detail in the section *4.2* of this paper.

It may be appropriate to consider the effect of the radiation from the SThM probe that may influence the experimental measurements. Previous works [4, 17, 25, 26] demonstrated that maximum heating is occurs at the end of cantilever. We can use average cantilever dimensions to estimate the radiation surface area [18, 19]:

From Stefan–Boltzmann law:

Here $P$ – power, radiated from surface area $S$, where $T$ – is the temperature of surface. This power for cantilever then is as follows:

$$P = ST^4 = 4 \times 10^{-10} \times [m^2] \, 5.7 \times 10^{-8} T^4 [Wm^{-2} K^{-4}] = 22.8 - 210^{-18} T^4 \, [W] \tag{9}$$

Heat flux to the end of cantilever is:

$$\Delta T \times 10^{-5} \, [WK^{-1}] \tag{10}$$

where $r$ is the typical thermal resistance of the cantilever of $10^{-5}$ WK$^{-1}$ [4, 6, 34]. Ratio between $P$ and $Q$ then will be:

$$\frac{Q}{P} = \frac{\Delta T}{rS\sigma T^4} \tag{11}$$

From these estimates it could be noticed that at most the temperature increase of the heater due to the radiation is by 0.01 K, allowing us to neglect heat radiation effects in our models.

*2.2 Numerical analysis of SThM measurements*

The previous works of SThM measurements demonstrated that considerable heat loss is occur at the end of cantilever [4, 24, 35]. For studying the influence of environment for SThM system sensitivity we consider two major type of geometries – similar to the analytical model, but more close to the reality of the conical SThM probe [24]. As stated above, the model here considers the zones II and III (as in Fig .2) of the SThM probe – the most critical parts defining the performance of SThM probe, these types are shown on Fig. 4.

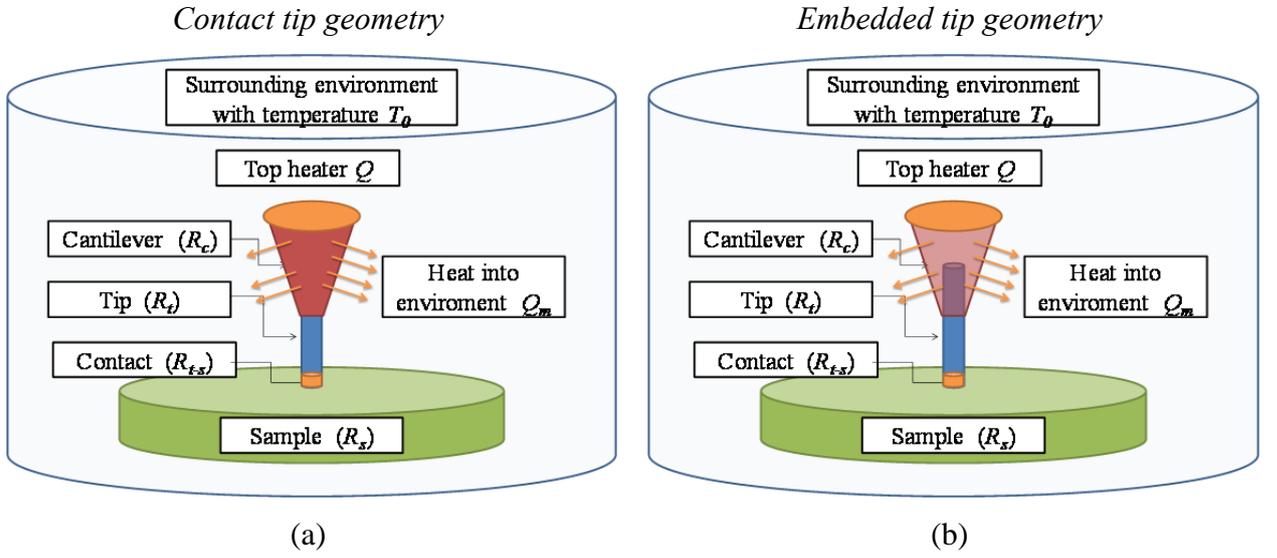

(a)            (b)

Fig 4. Types of modeling geometry (a) tip is in *contact* with cantilever, (b) tip is *embedded* into cantilever

It should be noticed that the modeling geometry has axial symmetry generally representing the very apex of the tip. It allows us to present heat equation in cylindrical coordinates:

$$\rho C_p \frac{\partial T}{\partial t} = \frac{1}{r}\frac{\partial}{\partial r}\left(kr\frac{\partial T}{\partial r}\right) + \frac{1}{r^2}\frac{\partial}{\partial \varphi}\left(k\frac{\partial T}{\partial \varphi}\right) + \frac{\partial}{\partial z}\left(k\frac{\partial T}{\partial z}\right) + g(r,\varphi,z,t) \qquad (12)$$

Here $T(r,\varphi,z,t)$ – is the temperature field, $g(r,\varphi,z,t)$ – density of internal heat generation, $\rho$, $k$ and $C_p$ – correspondingly, density, thermal conductivity and heat capacity of the material in the particular domain. We assume the temperature distribution is time independent, is symmetric about the z axis and without internal heat generation and the probe is heated from the top by the heater with the fixed heat $Q$. Simplifying the Eq. 12 we obtain:

$$Q = \frac{1}{r}\frac{\partial}{\partial r}\left(kr\frac{\partial T}{\partial r}\right) + \frac{\partial}{\partial z}\left(k\frac{\partial T}{\partial z}\right) \quad (13)$$

By solving Eq. 13 with boundary conditions for shapes we would consider in the model, we could obtain temperature distribution in our system including the temperature of the heater $T_h$. Ultimately, the ratio $T_h/Q$ is linked to the thermal resistance of the probe – a parameter measured in SThM.

Let's consider the main heat transfer pathways in this model subsystem where $Q$ is total heat source, $Q_m$ – heat lost to the environment, $Q_s$ – the heat transferred to the sample:

$$Q = Q_m + Q_s \quad (14)$$

Thermal resistance for SThM probe out of contact with the sample is:

$$\frac{Q}{T_{nc} - T_0} = \frac{1}{R_m} \quad (15)$$

here $T_{nc}$ – temperature at the top of tip when the tip is out of contact with sample, $T_0 = 300$ K is the temperature of the surrounding environment, $R_m$ – thermal resistance of environment (air, water, dodecane etc.). In contact, the Eq. 15 becomes:

$$\frac{Q}{T_{con} - T_0} = \frac{1}{R_m} + \frac{1}{R_c + R_t + R_{t-s} + R_s} \quad (16)$$

In order to include contact resistance in the FE simulation, we include in our model a thin resistive layer between the tip apex and the sample with height $h$ much less than the diameter of the contact and thermal conductivity calculated as:

$$k_{ts} = \frac{h}{R_{t-s}\pi r_t^2} = \frac{h}{\rho_{t-s}} \quad (17)$$

We use commercial finite element analysis (FE) package (COMSOL Multiphysics) to solve stationary heat equations and calculate temperature distributions and therefore thermal resistances in our system.

We calculated temperature distributions for two types of geometries – *contact* and *embedded* (Fig. 4) with tip length ranging from 0 nm to 1500 nm and investigating the effect of contact resistance on the heat distribution and materials used. Fig. 5 demonstrates example of temperature and heat flux distributions in the typical NW-SThM system.

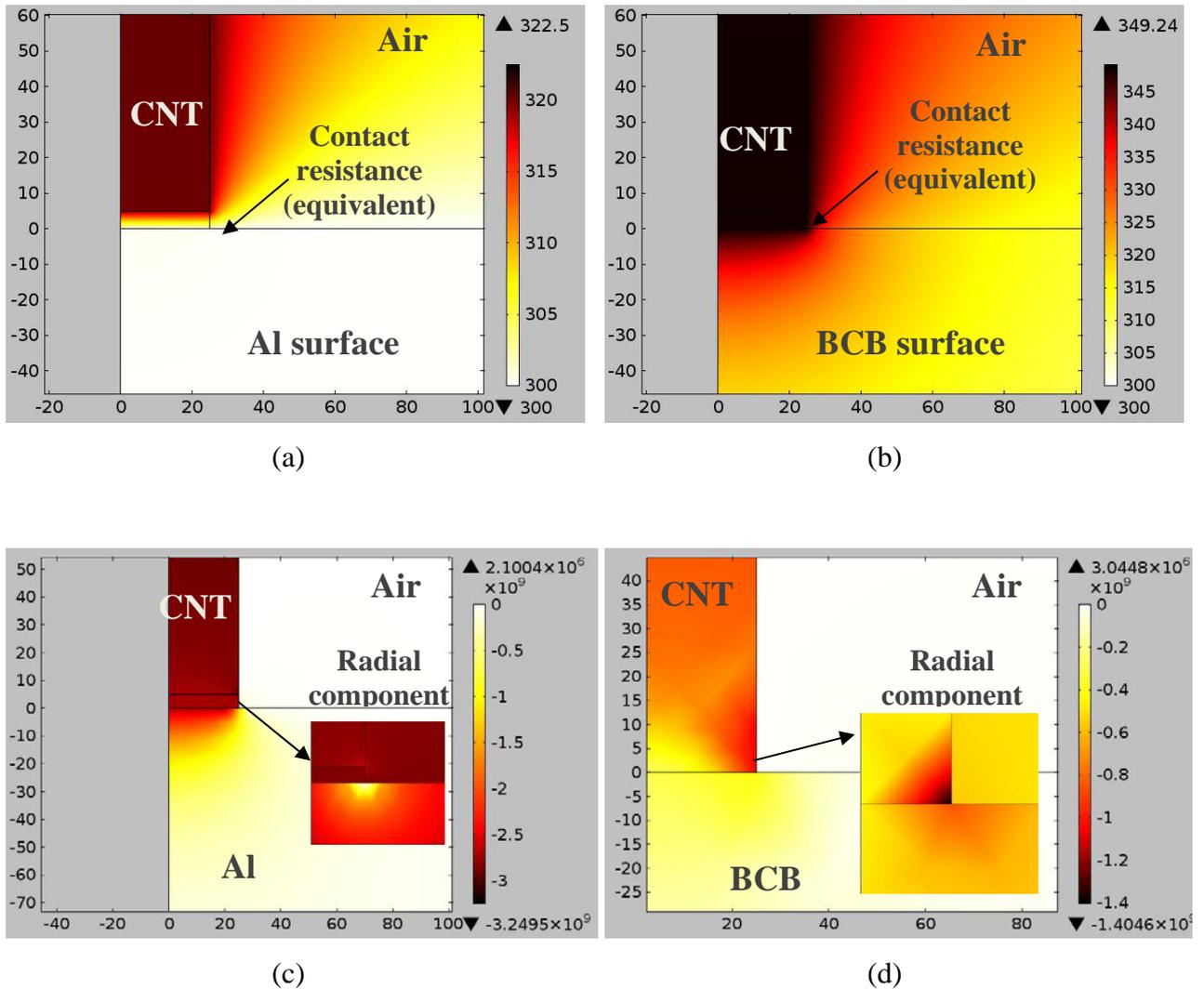

Fig. 5 (a) Cross section plot of a, b) temperature distribution and c) d) heat flux (z and r component) in the system with the SThM probe with 600 nm CNT - $Si_3N_4$ *embedded* probe. a, c) Al sample b, d) benzocyclobutene (BCB) sample; air environment.

Fig. 5 (a-d) illustrate the results of FE modeling of temperature and heat flux distribution in NW-SThM system with 600 nm CNT tip embedded in the $Si_3N_4$ probe for two substrate Al and BCB that have quite different thermal conductivities: $k_{BCB}$ = 0.29 Wm$^{-1}$K$^{-1}$ and $k_{Al}$ =237 Wm$^{-1}$K$^{-1}$ [36, 37]. Several qualitative features can be drawn from this simulation. First, the influence of contact resistance $\rho_{t-s}$ for Al sample is much more significant than for BCB sample, as the thermal resistance of the BCB sample itself is high in comparison with the corresponding contact thermal resistance. Also it should be noticed that for low thermal conductive BCB substrate heat flux from cantilever across the environment increase the effective radius of thermal contact (Fig. 5d with radial component of the heat transport notably high at the perimeter of the NW). In section *4* we apply this model for calculation of temperature distribution and thermal resistances for different geometries – *contact* and *embedded* as well as different lengths of CNT and GaAs NW-SThM probes ranging from 0 to 1000 nm, applied for measurements of Al, BCB, SiO$_2$ and graphene samples. Such model, if needed, can also be effectively applied to other material and environments systems.

## 3. Experimental methods

For experimental SThM measurements we used as standard Si$_3$N$_4$ SThM probes (Kelvin Nanotechnology) and same probes modified with CNT tips. CNTs attachment was described in details in [34] and Fig. 6 demonstrate scanning electron microscopy (SEM) image of SThM tip, modified by multiwall CNT:

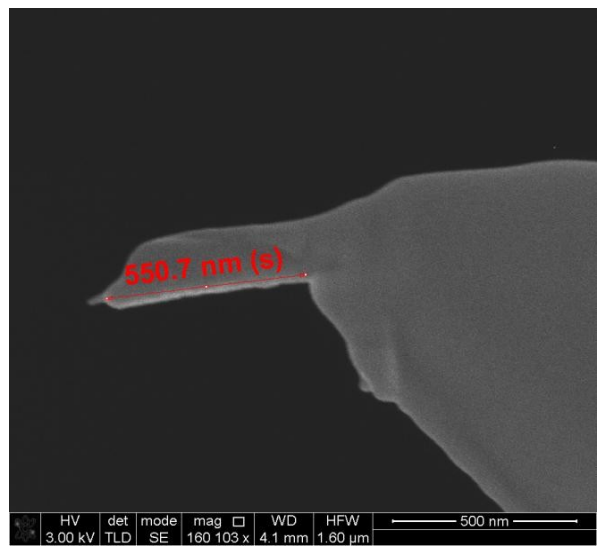

Fig. 6 SEM image of SThM probe modified with multiwall CNT tip, scale 500 nm.

The thermal calibration of the SThM probe was performed as described in [4]. The calibration allowed us to link probe resistance values with the probe temperature, and by knowing the power applied to the probe, to determine the absolute thermal resistance of the probe in and out of contact with the studied sample. The Al-BCB ultra large scale integrated polymer interconnects sample [38] was cleaned by sonication in acetone, isopropanol and DI water each for 10 minutes with final short plasma clean in an Ar/O$_2$ mix. As one of the challenges for current SThM probes is their low ability to discriminate between the samples with igh thermal conductivity, a sample of graphene on SiO$_2$ was used to test the CNT tips response to high conductive materials. Graphene has one of the highest known thermal conductivities in nature (on the order of 2000-5000 Wm$^{-1}$K$^{-1}$) [39, 40] hence why it was selected. The 280 nm SiO2 on Si wafer substrates for graphene samples deposition were cleaned similar to Al-BCB sample with graphene deposited by mechanical exfoliation using pressure sensitive tape [39]. The graphene flakes in this study were a 3 nm thick flake on a 280 nm thick SiO$_2$/Si wafer substrate.

## 4. Results and discussion.

*4.1 Comparison of FE simulation with experimental results.*

In SThM experiments the two well defined areas of two materials with different thermal conductive properties, such as Al-BCB and thin layers of graphene on Si/SiO$_2$ substrate were studied. Most convenient is scanning the sample surface in the perpendicular direction to the interface between materials that allows one to obtain the SThM response profiles for investigated structures and analyze the measured thermal resistances for each type of material - Al to BCB and graphene layer to SiO$_2$/Si substrate in our study. A ratio between two thermal resistances (Al to BCB or SiO$_2$ to graphene) gives us a measure of sensitivity of SThM to different materials. Two types of SThM probes were used – standard Si$_3$N$_4$ probe and same probe modified by MW-CNT. Analysis of results of two types SThM measurements allowed us to compare the sensitivity of each probe in the given system, as well as compare these results with FE simulations. According to generalized SThM model (Fig. 2) thermal resistance of cantilever and tip is:

$$\frac{Q}{(T_{con-i}-T_0)} - \frac{Q}{(T_{nc-i}-T_0)} = \frac{Q(T_{nc-i}-T_{con-i})}{(T_{con-i}-T_0)(T_{nc-i}-T_0)} = \frac{1}{R_h+R_t+R_{t-s}+R_s} = \frac{1}{R_i} \qquad (18)$$

where $T_{con-i}$ and $T_{nc-i}$ are, correspondingly, the temperatures of the probe in-contact and out-of-contact with the sample $i$ (Al, BCB, SiO$_2$ or graphene), and $R_i$ is defined as a thermal resistance of the probe in contact with material $i$. We will also define $R_0$ – the full thermal resistance of the *zone I* of SThM probe (Fig. 1, 2) that includes a heat flow to the cantilever base and thermal losses to the environment.

$$\frac{1}{R_0} = \frac{1}{R_{mfull}} + \frac{1}{R_c} \qquad (19)$$

Then experimentally measured in the SThM thermal resistance of the probe in contact with the sample $i$ or $j$ is:

$$\frac{1}{R_{i,j-m}} = \frac{1}{R_0} + \frac{1}{R_{i,j}} \qquad (20)$$

Using this description we can compare the experimentally measured values of thermal resistance ratios $R_{i-m} / R_{j-m}$, namely, $R_{Al-m} / R_{BCB-m}$ and $R_{GR-m} / R_{SiO2-m}$ in structures Al-BCB and graphene-on-SiO$_2$.

It should be noticed that one of the significant factors influencing SThM measurements at nanoscale range is contact thermal resistance of the interface between tip and substrate materials [22, 32, 41, 42]. While considering this thermal resistance is essential for correct modeling of SThM measurements, the literature estimates of such thermal resistance vary significantly and not always available for the particular studied materials. Therefore we used presented in [32] thermal resistances for different interfaces selecting pairs of materials which have the similar properties. Also, as we have seen in section *2*, the contact thermal resistance of BCB or SiO$_2$ is not important due to the high thermal resistance of the polymer itself, so it was neglected in this case. For Si$_3$N$_4$ probe - Al sample pair the similar pair is for MgO – Al [32], for Si$_3$N$_4$ – graphene – MgO-diamond, for CNT – Al-diamond-Al, and for CNT-graphene – diamond-diamond pair.

Fig. 8 demonstrates experimental results of SThM measurements of $1 - R_{i-m} / R_{j-m}$ for Al-BCB structures and results of modeling of this ratio depending on contact thermal resistance value ($\rho_{t-s}$). According [32] for C-Al interface $\rho_{C-Al} = 9\times10^{-9}$ Km$^2$W$^{-1}$ and for Si$_3$N$_4$-Al $\rho_{Si3N4-Al} = 4.5\times10^{-9}$ Km$^2$W$^{-1}$. The calculation was done using the models presented in Fig. 4 with cantilever apex length 1000 nm, top radius of conical apex – 500 nm, CNT tip total length – 1000 nm with CNT tip embedded into cantilever for 500 nm, CNT radius of 25 nm, the thermal conductivity of CNT is 1000 Wm$^{-1}$K$^{-1}$ [2] thermal conductivity of Al generally of 230 Wm$^{-1}$K$^{-1}$ [37], ambient temperature $T_0$ is 300 K, $Q_h$ selected such $T_{nc} \approx 360$ K, although it should be noted that $Q_h$ does not directly influence the measured thermal resistances.

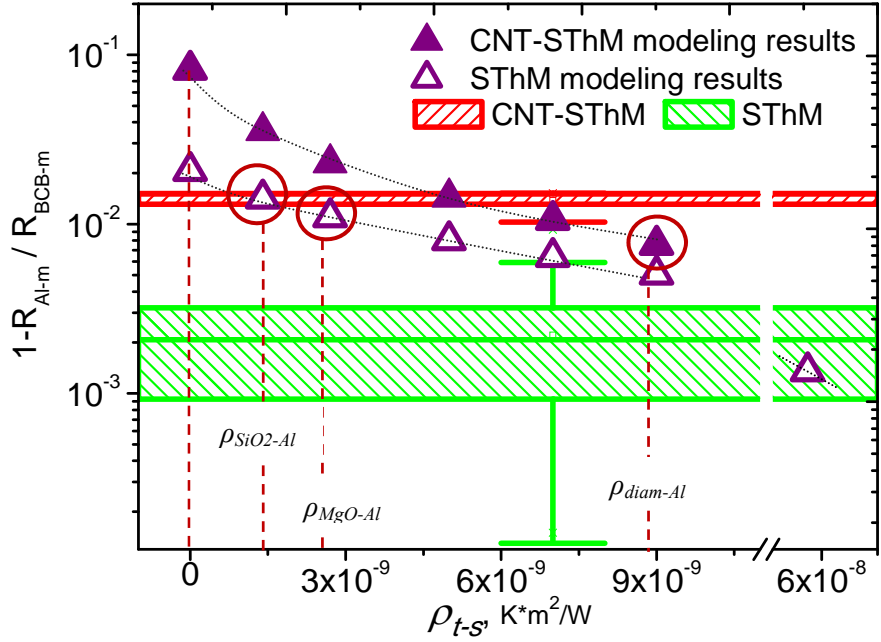

Fig. 8 Experimental results for ratio of $1-R_{Al-m}/R_{BCB-m}$ for measured thermal resistances of Al and BCB materials, green band - SThM experimental measurements for $Si_3N_4$ cantilever, red band – SThM experimental measurements for cantilever modified by CNT tip. Triangles – modeling results for different thermal contact resistances (horizontal axis) with open triangles corresponding to standard SThM probe and closed triangles for the CNT-SThM probe. Points corresponding to literature data [32] for contact resistances are in the red circles.

In Fig. 8 the green band correspond to the $Si_3N_4$ cantilever SThM experimental measurements of, $R_{Al-m}/R_{BCB-m}$ red band correspond to $Si_3N_4$ cantilever modified by CNT. The higher this ratio, the better is the SThM performance. We can clearly see that Fig. 8 demonstrates that cantilever modified by CNT tips improves the sensitivity of the SThM system [6, 34]. Also this figure clearly by comparing experimental results with FE modeling values for different contact resistance, we could obtain intersects between experiment data and results for FE simulated thermal resistances. This allows us to provide direct estimates of contact thermal resistances for interfaces of CNT-Al - $\rho_{t-s} = 5\times10^{-9} \pm 1\times10^{-9}$ $Km^2W^{-1}$ and for $Si_3N_4$-Al $\rho_{t-s} = 6\times10^{-8} \pm 2.5\times10^{-8}$ $Km^2W^{-1}$. It is interesting to note that the measured thermal contact resistance between CNT-Al is lower than between diamond and Al [32] of $\sim1\times10^{-8}$ $Km^2W^{-1}$ – that can be explained by the fact that multiwall CNT has lower Debye temperature compared with perfect diamond, as well as high anisotropy of CNT. It also looks that the contact between CNT and Al is relatively perfect, not leading to excess contact resistance. At the same time, the thermal resistance between standard $Si_3N_4$ SThM probe and Al is significantly higher than suggested by the literature, by the order of magnitude, suggesting that the contact between these materials may have multi-asperity nature [35] that can significantly degrade the contact resistance.

We also performed a similar comparison for experiments on graphene that has extremely high thermal conductance. Fig. 9 demonstrate comparison between experimental results of SThM measurements the $1 - R_{GR-m} / R_{SiO2-m}$ ratio between graphene flake and SiO$_2$ of Si and results of modeling calculations. First, one can notice that experimental results of SThM measurements for the Si$_3$N$_4$ cantilever and CNT modified cantilever have the same range of sensitivity value. It could indicate that the CNT cantilever is not significantly better for absolute values of thermal conductivity for graphene-SiO$_2$ SThM measurements, but, at the same time, the scatter of experimental results for CNT modified tips is much smaller than for Si$_3$N$_4$ cantilever that is a clear benefit. Secondly, while estimation of contact resistance values were based on values for interfaces between similar materials diamond-diamond and MgO-diamond, which are $7.3 \times 10^{-10}$ and $7.5 \times 10^{-10}$ Km$^2$W$^{-1}$ [32], experimentally measured ratios are significantly lower than predicted by FE in the absence of contact resistance. Comparison between experimental data with calculation results demonstrate that for interfaces CNT-graphene and Si$_3$N$_4$-graphene contact resistance will be significant large and close to $10^{-8}$ Km$^2$W$^{-1}$. This strongly suggests that there are other mechanisms leading to increased contact resistance that may be multi-asperity contacts, large anisotropy of graphene has in plane and normal to the graphene planes direction, as well as likely effects of the ballistic thermal conductance due to the large mean-free-path in both CNT and graphene layer [43].

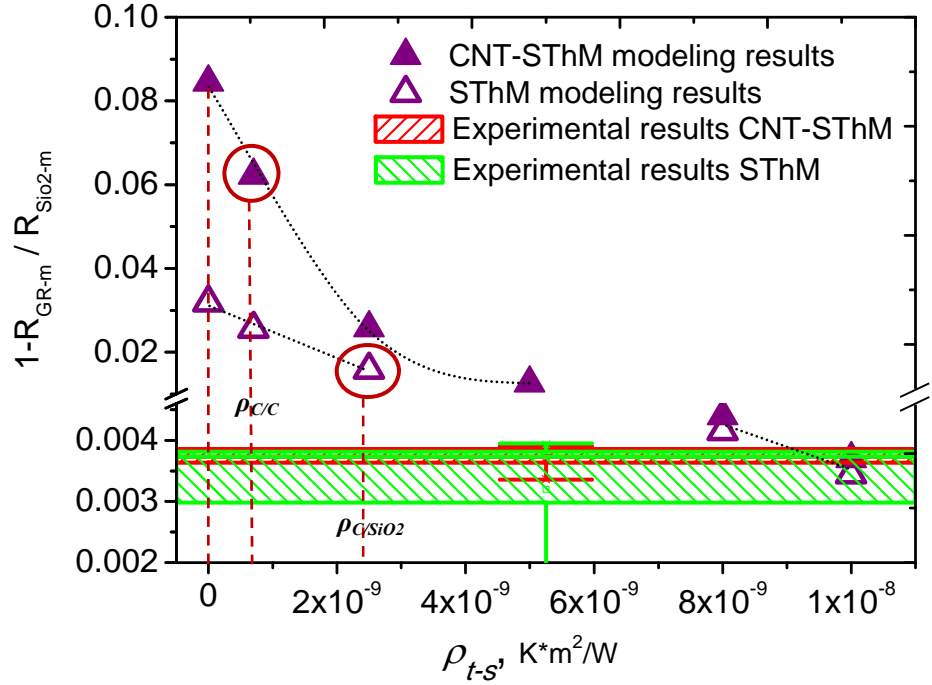

Fig. 9 Experimental results for ratio $1 - R_{GR-m}/R_{SiO2-m}$ for measured thermal resistance of graphene and SiO$_2$/Si, green band - Si$_3$N$_4$ probe, red band – CNT-SThM probe. Triangles - modeling results for different thermal contact resistances (horizontal axis) with open triangles corresponding to standard SThM probe and closed triangles for the CNT-SThM probe. Points corresponding to literature data [32] for contact resistances are in the red circles.

*4.2 Analytical and FEA models results*

In this work we developed analytical model of SThM cantilever which allows us to study contribution of different parameters into results of SThM measurements. According this model we could describe how the cantilever geometry, tip length, radius and material of tip and substrate will influence sensitivity of SThM. Also using a FEA analysis we studied a contribution of environment to the SThM measurements. Fig. 10 shows a comparison of the thermal resistance $R_i$ (6a) for *embedded* and *contact* geometries (as described in Fig. 3, 4) $R_{i\_emb}$ (6b) with the values of $R_i$ (18) obtained by numerical calculations. In Fig. 10 the dependences of thermal resistance on tip length and radius for contact and embedded geometries are presented. Lines correspond to analytical model and dots to numerical FE modelling. Parameters for numerical and analytical calculation are the same as in section *4.1*.

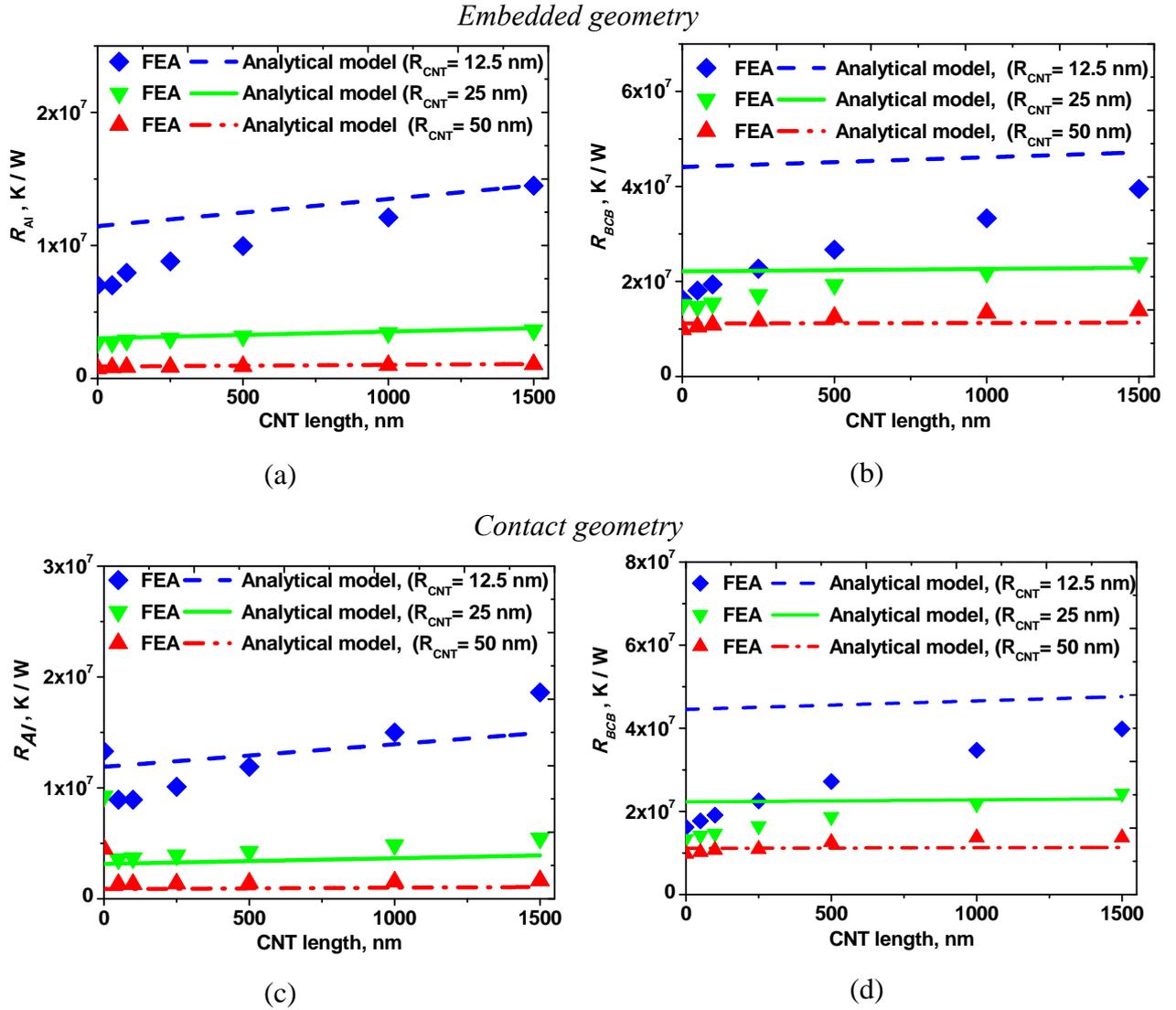

Fig. 10 Comparison of FE modeling (dots) and analytical estimates (lines) of sample thermal resistance $R_i$. Dependence of (a, c) Al and (b, d) BCB thermal resistance ($R_{Al}$) on CNT tip length for different CNT radii. a,b) *embedded* probe and c, d) – *contact* probe geometry.

Results for Al substrate shown on fig 10 (a) and (b) demonstrate a good agreement between numerical and analytical calculations. The differences in the values $R_i$ obtained in analytical and numerical models for BCB substrate, especially for small length of CNT, can be explained by an increase in the effective radius of heating in comparison with the real contact radius between CNT and BCB substrate due to the conductance through air (see Fig. 5).

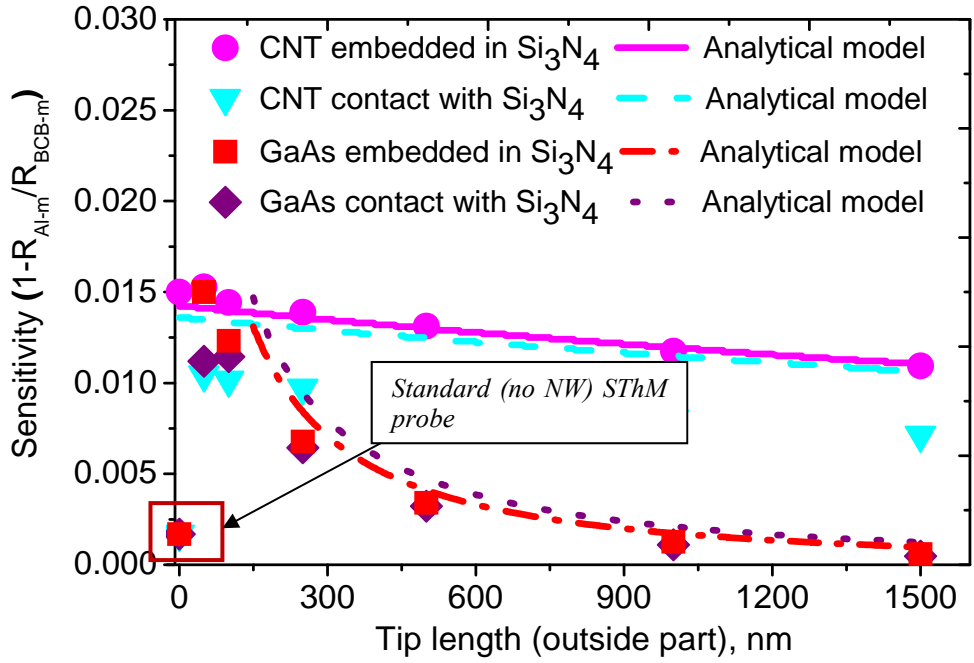

Fig. 11 Dependence of sensitivity SThM system on tip length for CNT tip and GaAs tip, dots – FEA calculation, lines –analytical model.

Analysis results presented on Fsig. 11 demonstrates that while modification of SThM probes with high thermal conductivity CNT does increase the sensitivity of SThM, the other NW tips – such as semiconducting GaAs can also improve the sensitivity of system if the NW is relatively short (100-200 nm) that may provide additional functionality to SThM.

## 5. Conclusions

In this paper we developed simple analytical model of SThM with a thermally conductive nanowire (NW) probe that allows elucidating the key phenomena in the SThM measurements of low and high thermal conductivity materials such as effects of NW thermal conductivity, geometrical parameters of the probe and NW, and the effects of contact resistance. The analytical model, corresponding FE modeling of standard SThM and NW-SThM probe and experimental data allowed us to provide direct estimates of contact thermal resistances for interfaces of CNT-Al of $5\times10^{-9} \pm 1\times10^{-9}$ $Km^2W^{-1}$ and for $Si_3N_4$-Al of $6\times10^{-8} \pm 2.5\times10^{-8}$ $Km^2W^{-1}$ suggesting that multi-asperity nature of the contact and anisotropy of the CNT may significantly influence the contact resistance. Our analysis also indicated that these models may be efficiently used for the NW with the radius of 25 nm and above, providing significant tools for the development of novel SThM probes that include semiconductor NW enabling additional functionalities in SThM.

## 5. Acknowledgments

Authors acknowledge useful discussion with Vladimir I. Falko from Lancaster University regarding the nature of nanoscale transport and physical interpretation of the thermal models. Authors also acknowledge input of Craig Prater and Roshan Shetty from Anasys Instruments for the support related to the SThM development. We acknowledge the support from the EPSRC grants EP/G015570/1 and EP/G017301/1, EP/K023373/1, EPSRC-NSF grant EP/G06556X/1 and EU FP7 grants, GRENADA (GA-246073) and FUNPROB (GA-269169), scientific programs of the Russian Academy of Sciences, grants of the Russian Foundation for Basic Research, contracts with the Russian Ministry of Education and Science, One of the authors B.A.D. gratefully acknowledges financial support of scholarship of the President of the Russian Federation for young scientists and graduate students.